\documentclass[lettersize,journal]{IEEEtran}
\usepackage{amsmath,amsfonts}

\usepackage{cite}
\usepackage{array}

\usepackage{tabularx}
\usepackage{booktabs}

\usepackage{bbding}
\usepackage[caption=false,font=normalsize,labelfont=sf,textfont=sf]{subfig}
\usepackage{multirow}
\usepackage{makecell}
\usepackage{textcomp}
\usepackage{stfloats}
\usepackage{url}
\usepackage{verbatim}
\usepackage{graphicx}
\usepackage[T1]{fontenc} 
\usepackage{amssymb} 
\usepackage{algorithmic}
\usepackage{algorithm}
\usepackage[backref=False]{hyperref}
\def\BibTeX{{\rm B\kern-.05em{\sc i\kern-.025em b}\kern-.08em
    T\kern-.1667em\lower.7ex\hbox{E}\kern-.125emX}}
\usepackage{balance}
\newcommand{\Rmnum}[1]{\uppercase\expandafter{\romannumeral #1}}  
\usepackage{tikz,xcolor,hyperref}

\usepackage{amsmath}
\usepackage{subcaption}
\usepackage{graphicx}
\usepackage{amssymb}

\definecolor{lime}{HTML}{A6CE39}
\DeclareRobustCommand{\orcidicon}{
	\begin{tikzpicture}
		\draw[lime, fill=lime] (0,0)
		circle[radius=0.16]
		node[white]{{\fontfamily{qag}\selectfont \tiny \.{I}D}}; 
	\end{tikzpicture}
	\hspace{-2mm}
}
\foreach \x in {A, ..., Z}{%
	\expandafter\xdef\csname orcid\x\endcsname{\noexpand\href{https://orcid.org/\csname orcidauthor\x\endcsname}{\noexpand\orcidicon}}
}

\begin{document}
\title{Adaptive Multi-Dimensional Coordinated Comprehensive Routing Scheme for IoV}

\author{Ruixing~Ren\hspace{-1.5mm}\orcidA{}, Minqi~Tao\hspace{-1.5mm}\orcidB{}, Junhui~Zhao\hspace{-1.5mm}\orcidC{},~\IEEEmembership{Senior~Member,~IEEE}, Qiuping~Li\hspace{-1.5mm}\orcidE{}, and Xiaoke~Sun\hspace{-1.5mm}\orcidD{}

\thanks{Corresponding author: Junhui Zhao.
		
Ruixing Ren, Junhui Zhao are with the School of Electronic and Information Engineering, Beijing Jiaotong University, Beijing 100044, China. (e-mail: renruixing0604@163.com; junhuizhao@hotmail.com)

Minqi Tao is with the School of Information and Software Engineering, East China Jiaotong University, Nanchang 330013, Chiina.
		
Qiuping Li and Xiaoke Sun are with the National Computer Network Emergency Response Technical Team/Coordination Center of China (CNCERT/CC), Beijing 100029, China.}
}

\maketitle

\begin{abstract}
  The characteristics of high-speed node movement and dynamic topology changes pose great challenges to the design of internet of vehicles (IoV) routing protocols. Existing schemes suffer from common problems such as insufficient adaptability and lack of global consideration, making it difficult to achieve a globally optimal balance between routing reliability, real-time performance and transmission efficiency. This paper proposes an adaptive multi-dimensional coordinated comprehensive routing scheme for IoV environments. A complete IoV system model including network topology, communication links, hierarchical congestion and transmission delay is first constructed, the routing problem is abstracted into a single-objective optimization model with multiple constraints, and a single-hop link comprehensive routing metric integrating link reliability, node local load, network global congestion and link stability is defined. Second, an intelligent transmission switching mechanism is designed: candidate nodes are screened through dual criteria of connectivity and progressiveness, a dual decision-making of primary and backup paths and a threshold switching strategy are introduced to avoid link interruption and congestion, and an adaptive update function is constructed to dynamically adjust weight coefficients and switching thresholds to adapt to changes in network status. Simulation results show that the proposed scheme can effectively adapt to the high dynamic topology and network congestion characteristics of IoV, perform excellently in key indicators such as routing interruption times, packet delivery rate and end-to-end delay, and its comprehensive performance is significantly superior to traditional routing schemes.
\end{abstract}


\section{Introduction}
With the rapid development of intelligent transportation systems, the internet of vehicles (IoV), as a core enabling technology, has become a research hotspot in both academia and industry \cite{RenIoV}. By enabling low-latency and high-reliability communications between vehicle-to-vehicle (V2V) and vehicle-to-infrastructure (V2I), IoV provides a critical data interaction platform for applications such as autonomous driving, intelligent route planning, and traffic safety early warning \cite{Castillo,RenITS}.

However, IoV networks exhibit distinct characteristics, including high-speed node mobility, dynamically changing topology, drastic channel quality variation, and frequent network congestion. These pose severe challenges to designing efficient and robust routing protocols \cite{Cunha,RenRIS}. How to guarantee high packet delivery ratio (PDR), low end-to-end delay, and low bit error rate (BER) in such complex and dynamic environments remains a core urgent issue in current IoV routing research.

Extensive research has been conducted on routing optimization in IoV environments. Early studies mainly focused on topology-based routing protocols, but they incur enormous routing overhead and are prone to link breakage in highly dynamic IoV scenarios \cite{10951108}. Recently, Wang et al. \cite{Wang} integrated a dynamic topology evolution model with routing optimization to provide a basis for topology prediction in decision-making. However, the coordination mechanism between local link changes and global topology optimization was not clearly reflected, leaving room for improvement in multi-objective performance balance.

To improve protocol scalability, geographic location-based routing protocols have emerged \cite{Geo}. However, it is prone to local optimality in the case of network holes or uneven node distribution. Ye et al. \cite{Ye} used satellite maps and SD maps to predict lane maps, alleviating the local optimality problem caused by over-reliance on local perception. Suo et al. \cite{Suo} utilized node location information to construct proactive multipath routes, and adopted adaptive path adjustment as well as a static-node-based local routing maintenance scheme. Nevertheless, end-to-end connected paths may be absent for a long time in scenarios such as sparse vehicle density or suburban areas. To address this, opportunistic routing or delay-tolerant network routing ideas have been introduced \cite{11397273}, abandoning the pre-setting of complete paths and adopting a store-carry-forward mode \cite{11237070}. This mode uses vehicle mobility as a transmission medium, temporarily storing data when no suitable next hop is available until a forwarding opportunity arises.

With the enrichment and differentiation of IoV application scenarios, a single routing strategy can hardly meet all requirements, making hybrid and clustering routing protocols important research directions. Hybrid routing addresses complex and dynamic network environments by integrating the advantages of different routing strategies \cite{Atlantis}. For example, geography-based unicast routing is adopted in high-vehicle-density areas, switching to opportunistic routing in sparse areas \cite{Geo_Routing}; alternatively, cluster-structure routing is used at the backbone network level, while simple flooding or geographic forwarding is employed within clusters. Clustering routing forms a virtual backbone network through dynamic vehicle clustering and cluster head election, reducing nodes involved in routing calculation, lowering control overhead, and improving routing scalability and stability \cite{Bi}.

In recent years, the development of artificial intelligence has profoundly reshaped the research paradigm of IoV \cite{Yao,RenUAV}. In particular, intelligent routing algorithms based on deep reinforcement learning (DRL) have emerged as a cutting-edge research topic. Moon et al. \cite{Moon} took the dynamic traffic environment as the state input and achieved dynamic routing decision-making through agent training.
By enabling continuous interaction between routing entities and the environment, such methods learn optimal forwarding policies under various network states from historical or real-time data, without explicitly predefined complex routing rules. They thus exhibit great potential in handling high-dimensional state spaces and complex optimization objectives, and can adaptively learn and approximate optimal routing strategies in dynamically uncertain environments \cite{6G_V2X}. For instance, a DRL-based efficient routing algorithm for intelligent QoS optimization selects optimal data transmission paths by dynamically adapting to vehicular network variations \cite{Shitong}.
On this basis, more studies have proposed targeted DRL routing schemes to address the unique characteristics of IoV, such as highly dynamic topology, unpredictable contact patterns, and resource constraints. As an example, the CR-DRL algorithm, which integrates an actor-critic framework and heuristic functions, realizes real-time optimal relay selection and dynamic overlapping cluster adjustment \cite{11397273}.

Although existing IoV routing protocols have achieved performance optimization in different scenarios, they still have common shortcomings. Topology-based protocols fail to adapt to highly dynamic topologies. Geography-based protocols tend to fall into local optima. Opportunistic routing is only suitable for sparse scenarios and cannot balance transmission efficiency in dense ones. Cluster management and strategy switching mechanisms of hybrid and clustering routing remain to be improved. While DRL-based routing algorithms exhibit certain adaptability, they mostly focus on local network states, neglect global congestion and link stability, and rely on fixed parameters that are difficult to adapt to dynamic IoV scenarios, thus failing to globally optimize routing reliability, latency and transmission efficiency.

To address the above issues, this paper proposes an adaptive multi-dimensional coordinated comprehensive routing scheme for IoV. The main contributions are summarized as follows:
\begin{itemize}
	\item We model the IoV system and establish a multi-dimensional comprehensive evaluation model that incorporates link reliability, local node load, global network congestion, and link stability, providing a comprehensive and objective quantitative basis for routing decisions.
	\item An intelligent V2I/V2V handover mechanism is designed. Candidate nodes are selected via connectivity and progressiveness, while primary-backup dual-path and threshold-based switching strategies are adopted to avoid link disruptions and congestion. An adaptive update function is developed for diverse network densities and loads, enabling real-time adaptation to network dynamics.
	\item Simulation results demonstrate that the proposed scheme outperforms conventional approaches in key metrics including routing disruption frequency, packet delivery ratio, and end-to-end delay, verifying its effectiveness and superiority in highly dynamic IoV environments.
\end{itemize}

The rest of this paper is organized as follows. Section \ref{2} constructs the IoV system model and formulates the problem. Section \ref{3} presents the proposed scheme. Section \ref{4} validates the proposed scheme through simulation experiments. Section \ref{5} concludes the paper.

\begin{figure}[t]
	\centerline{\includegraphics[width=3.5in,keepaspectratio]{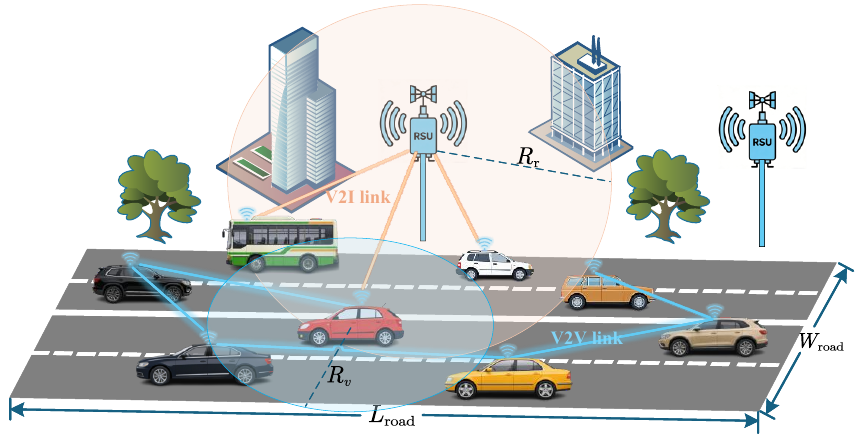}}
	\caption{IoV System Network Topology Model.}
	\label{fig1}
\end{figure}
\section{System Model and Problem Formulation}\label{2}
\subsection{Network Topology Model}
The IoV system constructed in this paper consists of two communication entities: vehicle nodes and road side units (RSUs). As shown in Fig. \ref{fig1}, the road region is defined as a 2D rectangular space $\mathcal{R}=\left [ 0,L_{\mathrm{road}} \right ] \times \left [ 0,W_{\mathrm{road}}  \right ] $. The set of vehicle nodes is denoted by $\mathcal{V}=\left \{ v_1,v_2,...,v_n,...,v_N \right \} $, where $N$ is the total number of vehicles. Each vehicle $v_n$ has a communication radius $R_v$, and its state is jointly described by the position coordinate $\mathbf{P}_n(t)=\left[ x_n(t),y_n(t)\right]^T$, velocity vector $\mathbf{v}_n(t)=\left[ v_{nx}(t),v_{ny}(t)\right]^T$, and queue load $q_n(t)$. The vehicle mobility model satisfies the following:
\begin{equation}
	\left\{\begin{matrix}
		x_n(t+\Delta t)=x_n(t)+v_{nx}(t)\cdot \Delta t \\
		y_n(t+\Delta t)=y_n(t)+v_{ny}(t)\cdot \Delta t
	\end{matrix}\right.
\end{equation}
where $\Delta t$ is the time slot interval. The set of RSU nodes is denoted as $\mathcal{R}=\left \{ r_1,r_2,...,r_m,...,r_M \right \} $, where $M$ is the total number of deployed RSUs. Each RSU has a much larger communication coverage than vehicle nodes, i.e., $R_r > R_v$. Each $r_m$ can establish direct communication links with all vehicle nodes within its communication coverage, which assists in forwarding data packets and alleviates the link instability and network congestion of V2V communications.

\subsection{Communication Link Model}
The communication link set of the IoV can be expressed as the union of two core modes: $L=L_{\mathrm{V2V}}\cup L_{\mathrm{V2I}}$. The V2V direct link set is defined as
\begin{equation}
	L_{\mathrm{V2V}}=\left \{ l(v_i,v_j):v_i,v_j\in\mathcal{V},d(v_i,v_j)\le R_v \right \}.
\end{equation}
The condition is that the Euclidean distance between two vehicle nodes $v_i$ and $v_j$ does not exceed the vehicle communication radius $R_v$. The V2I communication link set is defined as
\begin{equation}
	L_{\mathrm{V2I}}=\left \{ l(v_n,r_m):v_n\in\mathcal{V}, r_m\in \mathcal{R}, d(v_n,r_m)\le R_r \right \}.
\end{equation}
This holds if the Euclidean distance between vehicle node $v_n$ and RSU node $r_m$ does not exceed the communication radius of the RSU.

In urban IoV scenarios, signal propagation is affected by path loss, shadow fading, multipath fading, adjacent-channel interference, and other factors. The channel model established in this paper comprehensively considers the above factors. Based on the received power $P_r$, single-hop link quality is characterized from three dimensions: signal strength, transmission error, and data reception success rate, using three core metrics: signal-to-noise ratio (SNR), BER, and packet reception rate (PRR).

For any single-hop communication link $l(n_i,n_j)\in L$ in the IoV, where $n_i,n_j\in \mathcal{V}\cup\mathcal{R}$, the received power $P_r$ at the receiver \cite{RenDCAN} adopts the unified form as follows
\begin{equation}
	P_r(d(n_i,n_j))=P_t \cdot G_t \cdot G_r \cdot (\frac{\lambda}{4\pi d_0})^2 \cdot (\frac{d_0}{d(n_i,n_j)})^\zeta \cdot X_\sigma \cdot \left | h \right |^2 
\end{equation}
where $P_t$ is the transmit power of the transmitting node, $G_t$ and $G_r$ are the transmit and receive antenna gains, respectively, $\lambda$ is the carrier wavelength, $d_0$ is the reference distance, $\zeta$ is the path loss exponent, and $X_{\sigma}$ is the shadow fading factor following a log-normal distribution $10\log_{10}X_\sigma \sim \mathcal{N}(0,\sigma^2)$, which characterizes the slow fading caused by fixed obstacles. $\left| h \right |^2 $ is the amplitude of multipath fading, characterizing the fast fading after signal reflection and scattering.

The SNR is defined as the ratio of the useful signal power to the noise power at the receiver:
\begin{equation}
	\mathrm{SNR}(n_i,n_j)=\frac{P_r(d(n_i,n_j))}{N_0}
\end{equation}
A higher SNR indicates stronger anti-interference capability of the link. The BER is approximated by the complementary error function:
\begin{equation}
	\mathrm{BER}(n_i,n_j)\approx \frac{1}{2}\mathrm{erfc}(\sqrt{\frac{\mathrm{SNR}(n_i,n_j)}{2}})
\end{equation}
It directly reflects the accuracy of data transmission.

PRR is a core metric characterizing the link data reception success rate, defined as the ratio of the number of data packets successfully received by the receiver to the total number transmitted by the sender. It comprehensively reflects the overall link transmission quality and serves as a key basis for link selection in routing algorithms. Under ideal assumptions (independent bit errors, no channel coding or interleaving), PRR and BER satisfy the exponential relationship:
\begin{equation}
	\mathrm{PRR}(n_i,n_j)=(1-\mathrm{BER}(n_i,n_j))^{L_p}
\end{equation}
where $L_P$ is the packet length. However, in urban vehicular communication scenarios, the application of channel coding and interleaving alters the mapping between BER and PRR, such that they no longer follow the simple exponential relationship under ideal assumptions. Meanwhile, as BER→0, PRR→1; as BER→1, PRR→0. Such extreme values cause singularity in routing metrics, impairing the numerical stability and decision rationality of the algorithm. Therefore, to simplify computation, avoid extreme-value interference, and ensure accuracy suitable for short-range urban vehicular communication, this paper adopts the following clipped approximation:
\begin{equation}
	\mathrm{PRR}(n_i,n_j)=\mathrm{clip}(1.0-\mathrm{BER}(n_i,n_j),0.01,0.999)
\end{equation}
where $\mathrm{clip}(\cdot)$ denotes the clipping function, which strictly confines PRR to the interval $\left [ 0.01,0.999 \right ] $.This avoids extreme-value interference while maintaining sufficient accuracy in short-range vehicular communication, providing stable quantitative support for subsequent routing decisions.

\subsection{Congestion Model}
To accurately capture congestion dynamics, this paper establishes a hierarchical congestion model from node-level and global-level dimensions, providing a quantitative basis for routing decisions.

Node-level congestion focuses on the local load state of a single vehicle node. This paper adopts normalized node load as the core metric.
The node load is dynamically updated with packet forwarding during operation and decays naturally over time.
To eliminate the deviation caused by differences in buffer capacity among nodes, the normalized queue load is defined as
\begin{equation}
	q(v_n)=\min(1,\frac{L_n(t)}{L_{\max}})\in[0,1]
\end{equation}
where $L_n(t)$ is the current actual queue length, and $L_{\max}$ is the maximum buffer capacity of the node. The value of $q(v_n)$ is positively correlated with the node congestion level, which can be broadcast to one-hop neighbors via beacons to provide a basis for routing selection.

Node-level congestion metric only reflects the local load of a single node, and cannot characterize global congestion caused by excessive vehicle density in the region and over-occupied channel resources. This paper selects channel occupancy as the global congestion metric. Assuming vehicles are uniformly distributed, the vehicle density in the region is $N/(W_{\mathrm{road}}\cdot L_{\mathrm{road}})$. The communication coverage area of each vehicle is $\pi R_v^2$, but the coverage areas of different vehicles overlap.
The channel occupancy level can be approximated as the ratio of the total coverage area of all vehicles to the total area of the region \cite{GARCIAVIDAL2025100978}. The formula for the global congestion level $C_\mathrm{global}$ is given as follows:
\begin{equation}
	C_{\mathrm{global}}=\min(1,\frac{N\cdot\pi R_v^2}{W_{\mathrm{road}}\cdot L_{\mathrm{road}}})
\end{equation}
Its value is positively correlated with the degree of global congestion: a higher overlap indicates greater vehicle density in the region, more intense channel contention, and more severe global congestion.

\subsection{Delay Model}
The end-to-end transmission delay in IoV consists of three core components, corresponding to the three physical processes: signal propagation, data transmission, and queue waiting. In the short-range communication scenario of IoV, the propagation delay is usually small and can be neglected.

Transmission delay is the time required to send a fixed-size data packet from node $n_i$ to node $n_j$, which is related to packet length and link transmission rate:
\begin{equation}
	\tau^t(n_i,n_j)=\frac{L_p}{\mathrm{rate}(n_i,n_j)}, n_i,n_j\in\mathcal{V}\cup\mathcal{R}
\end{equation}
where $\mathrm{rate}(n_i,n_j)$ is the physical-layer transmission rate of link $(n_i,n_j)$, determined by link distance and channel conditions. Vehicle nodes use normalized load to approximate queuing delay, adapting to distributed, low-overhead routing decisions:
\begin{equation}
	\tau_V^q(v_n)=\tau_0\cdot q(v_n), v_n\in\mathcal{V}
\end{equation}
where $\tau_0$ is the maximum queuing delay reference, used to map the load to a reasonable delay magnitude. As a centralized infrastructure, the RSU uses an in-degree-processing time model to characterize queuing delay, reflecting its scheduling capability and processing pressure:
\begin{equation}
	\tau_R^q(r_m)=k_R \cdot \deg_{\mathrm{in}}(r_m,t) \cdot \tau_R, r_m \in \mathcal{R}
\end{equation}
where $k_R$ is the RSU scheduling coefficient, used to adjust the impact of scheduling efficiency on delay \cite{app13074329}. $\mathrm{deg}_{\mathrm{in}}(r_m,t)$ is the number of active incoming links to RSU $r_m$ at time $t$. $\tau_R$ is the average processing time of a single data packet for the RSU.

The available path $\mathcal{P}$ from source node $n_S$ to destination node $n_D$ is defined as an ordered sequence consisting of $k+1$ nodes:
\begin{equation}
	\mathcal{P}=\left \{ n_S=n_0,n_1,n_2,...,n_D=n_k \right \}, \forall n_i\in\mathcal{V}\cup\mathcal{R} 
\end{equation}
where any adjacent node pair $(n_i,n_{i+1})$ must satisfy $d(n_i,n_{i+1})\le R_v $ or $R_r$, i.e., a valid direct communication link exists. The path must also satisfy the acyclic constraint, i.e., contain no repeated nodes, to avoid routing loops and the resulting resource waste and delay surge. The total end-to-end delay $\tau_{\mathrm{total}}$ of the transmission path $\mathcal{P}$ from the source node to the destination node can be expressed as:
\begin{equation}
	\begin{aligned}
		\tau_{\mathrm{total}}&= \sum_{n_i,n_j\in \mathcal{P}}\tau^t(n_i,n_j) \\&+ \sum_{v_n\in\mathcal{P},v_n\ne D}\tau^q_V(v_n) + \sum_{r_m\in\mathcal{R},r_m\ne D}\tau^q_R(r_m)
	\end{aligned}
\end{equation}

\subsection{Routing Problem Formulation}
The total end-to-end routing cost of path $\mathcal{P}$ is defined as the sum of the comprehensive routing metrics of each single-hop link:
\begin{equation}
	\mathrm{Cost}(\mathcal{P})=\sum_{i=1}^{k-1}M(n_i,n_{i+1})
\end{equation}
where $M(n_i,n_{i+1})$ is the comprehensive routing metric for a single-hop link, which jointly reflects link quality, node load, network congestion, and path stability.

The core objective of routing optimization is to select the optimal path $\mathcal{P}^\ast$ with the minimum total cost from all available paths $\Psi $, while ensuring routing validity and quality of service:
\begin{align*}
	\mathcal{P} 1: \mathcal{P}^\ast = \;&\mathrm{arg}\min_{\mathcal{P}\in\Psi} \mathrm{Cost}(\mathcal{P})
	\\
	s.t. \;\;  &\mathrm{C}1: d(n_i,n_j)\le R_v, \nonumber
	\\
	\,\,   &\mathrm{C}2: d(n_{i+1},n_D)<d(n_i,n_D), n_{i+1}\ne n_D,\nonumber
	\\
	\,\,   &\mathrm{C}3: \mathrm{PDR}(\mathcal{P})=\prod\mathrm{PRR}(n_i,n_j) \ge \mathrm{PDR}_{\min}, \nonumber
	\\
	\,\,   &\mathrm{C}4: q_n \le q_{\max}, v_n \ne n_D, \nonumber
	\\
	\,\,   &\mathrm{C}5: \tau_{\mathrm{total}}\le T_{\max}, \nonumber
\end{align*}
C1 ensures physical connectivity of the path, i.e., each hop must be within communication range. C2 ensures each hop is closer to the destination node $n_D$ than the previous hop. C3 guarantees path reliability and avoids severe packet loss caused by poor link quality, where PDR is the end-to-end packet delivery ratio. C4 prevents overloaded nodes, since packet dropping may occur if the queue of an intermediate node is full. C5 is the end-to-end total delay constraint.

\section{Proposed Scheme}\label{3}
The previous section has abstracted the IoV routing problem as a single-objective optimization model with multiple constraints, clarifying the core goal of finding the path with minimum total routing cost while ensuring constraints such as path connectivity, reliability, congestion control, and end-to-end delay. According to this objective, the single-hop comprehensive routing metric $M(n_i,n_{i+1})$ should adopt multi-dimensional criteria, including link reliability, node load, global congestion, and path stability.

Link stability quantifies the resilience of a link to topology changes by combining relative speed and communication distance:
\begin{equation}
	\begin{aligned}
		s(n_i,n_{i+1})&=1-0.5\times \min(1,\frac{v_{\mathrm{rel}}(n_i,n_{i+1})}{v_{\mathrm{ref}}})\\ 
		&-0.5\times\min(1, \frac{d(n_i,n_{i+1})}{R_v})
	\end{aligned}
\end{equation}
where $v_{\mathrm{rel}}(n_i,n_{i+1})$ is the relative speed between nodes $n_i$ and $n_{i+1}$, and $v_{\mathrm{ref}}$ is the reference speed threshold. A larger value of $s(n_i,n_{i+1})\in(0,1]$ indicates a more stable link. Then $M(n_i,n_{i+1})$ can be obtained as:
\begin{equation}
	\begin{aligned}
		M(n_i,n_{i+1})&=\alpha\cdot\frac{1}{\mathrm{PRR}(n_i,n_{i+1})} + \beta\cdot q_{n_{i+1}} \\&+ \gamma \cdot C_{\mathrm{global}} + \delta \cdot (1-s(n_i,n_{i+1}))
	\end{aligned}
\end{equation}
where $\alpha,\beta,\gamma,\delta$ are weighting coefficients satisfying $\alpha+\beta+\gamma+\delta=1$.

\begin{algorithm}[!t]
	\caption{The designed routing selection algorithm}
	\label{alg:ca_routing}
	\begin{algorithmic}[1]
		\REQUIRE $n_S$ $n_D$, $\mathcal{V}$, $\mathcal{R}$, $C_{th}$, etc.
		\STATE Calculate the global congestion degree $C_{\text{global}}$ by Eq. (10).\\
		\STATE It is defaulted to the V2V mode.
		\FOR{Each $r_m\in\mathcal{R}$}
		\IF{The source node can communicate with this RSU.}
		\STATE Switch to V2I mode; \textbf{break}
		\ENDIF
		\ENDFOR
		
		\IF{The mode is V2I.}
		\FOR{Each reachable RSU and each vehicle node}
		\IF{The vehicle satisfies both V2I connectivity and the destination-oriented advancement constraint}
		\STATE Add the vehicle to the candidate set $\Psi$
		\ENDIF
		\ENDFOR
		\ENDIF
		
		\IF{$\Psi=\emptyset$}
		\FOR{Each vehicle node}
		\IF{The vehicle satisfies both V2V connectivity and the destination-oriented advancement constraint.}
		\STATE Add the vehicle to the candidate set $\Psi$.
		\ENDIF
		\ENDFOR
		\ENDIF
		
		\FOR{Each vehicle node $\in\Psi$}
		\STATE Calculate the PRR (Eq. (8)), stability (Eq. (17)), load (Eq. (9)), and $M(n_i,n_{i+1})$ (Eq. (18)).
		\IF{The current value of $M(n_i,n_{i+1})$ is smaller}
		\STATE Update the primary path to this node.
		\ENDIF
		\IF{The current stability is higher.}
		\STATE Update the backup path to this node.
		\ENDIF
		\ENDFOR
		
		\STATE next hop = primary path
		\IF{The primary path is valid and its metric exceeds the threshold, and the backup path is valid and different.}
		\STATE next hop = backup path
		\ENDIF
		\RETURN next hop
	\end{algorithmic}
\end{algorithm}

Based on the above multi-dimensional metric, to further address the core challenges of IoV—high dynamics, frequent congestion, and time-varying topology—this paper proposes three key technical strategies. These strategies form a complete optimal routing selection framework from the perspectives of robustness enhancement, resource optimization, and scenario adaptation. The routing decision procedure of the proposed scheme is shown in Algorithm \ref{alg:ca_routing}.

(1) Global congestion state evaluation. Node $n_i$ first obtains the current global congestion level $C_\mathrm{global}$. If it is within the coverage of an RSU, the value can be obtained from the periodic broadcast of the global congestion statistics by the RSU. Otherwise, the node can estimate it by counting the local average number of neighbors.

(2) Adaptive V2I/V2V transmission mode selection. By continuously monitoring beacon messages broadcast by the RSU, the node determines whether it is within RSU coverage. If an RSU with signal strength above a predefined threshold is detected, the V2I mode is preferred. If no available RSU is detected, the node switches to V2I mode.

(3) Valid candidate node screening. According to the current transmission mode, neighbor nodes are subjected to dual screening (satisfying constraints C1: connectivity, C2: advancement):
\begin{itemize}
	\item V2I mode: Select vehicle nodes that communicate with both the reachable RSU and the source node $n_i$, and are closer to the destination node $n_D$.
	\item V2V mode: Select vehicle nodes within the communication range of the source node and closer to the destination node $n_D$.
	\item If the candidate set is empty after screening, enter the recovery mode, temporarily relax the geographic advancement constraint, or adopt the carry-and-forward strategy.
\end{itemize}

(4) Primary and backup path decision-making. For each valid candidate node $n_j$, the algorithm calculates the link stability $s(n_i,n_j)$ and comprehensive metric $M(n_i,n_j)$ by Eq. (17) and Eq. (18), and determines the next hops of the primary and backup paths based on these values. The next hop of the primary path is selected as the node with the minimum $M(n_i,n_j)$, representing the optimal comprehensive routing quality. The next hop of the backup path is chosen as the node with the maximum $s(n_i,n_j)$, representing the most stable link. When the comprehensive metric $M$ of the primary path exceeds the predefined threshold $C_{\mathrm{th}}$, the algorithm automatically switches to the backup path to avoid potential link breakage or severe congestion, thus improving routing robustness. Finally, the selected next-hop forwarding node is output.

Notably, the performance of the algorithm highly depends on the selection of the weighting coefficients $\alpha,\beta,\gamma,\delta$ and the switching threshold $C_{\mathrm{th}}$. Different network scenarios exhibit different sensitivities to each metric, and static parameters cannot achieve optimality across all scenarios. Therefore, we design the following adaptive adjustment strategy.

When the network density is high (i.e., $C_{\mathrm{global}}$ is large), global congestion becomes the dominant issue. It is necessary to increase $\gamma$ to enhance congestion avoidance, while appropriately reducing $\alpha$ and $\delta$. When the network density is low, the risk of link disruption increases, and $\delta$ should be enlarged to emphasize stability. The dynamic weighting function is designed as follows:
\begin{equation}
	\gamma=\gamma_0 \cdot (1+C_{\mathrm{global}}),
\end{equation}  
\begin{equation}
	\delta=\delta_0 \cdot (1-C_{\mathrm{global}})
\end{equation}
where $\gamma_0$ and $\delta_0$ are the initial baseline values. When the queue length of node $v_n$ is high, it indicates local congestion. Thus, $\beta$ should be increased to steer traffic away from other congested nodes and avoid aggravating congestion. The formula is adopted as follows:
\begin{equation}
	\beta=\beta_0 \cdot (1+q(v_n))	
\end{equation}
These weights are then normalized together with the other weights, so that heavily loaded nodes prefer to select next-hop nodes with shorter queues. When frequent link disruptions are detected, $C_{\mathrm{th}}$ should be appropriately decreased to improve switching sensitivity and reduce disruptions. Conversely, if excessive switching causes routing oscillations, $C_{\mathrm{th}}$ is increased to stabilize the routing.

\section{Simulation Results and Analysis} \label{4}
In this section, simulation experiments are conducted to verify the effectiveness of the proposed scheme. The initial $x$ and $y$ coordinates of vehicles follow a uniform distribution. The vehicle speed follows a uniform distribution over 15$\sim$30 m/s. The vehicle heading angle follows a uniform distribution over -0.05$\sim$0.05 rad. The number of RSUs $M$ is given by $\max(5, N // 200)$, and RSUs are uniformly and equidistantly deployed along the road centerline. The maximum hop count is set to 15.

The mainstream 2.4 GHz wireless communication band for IoV is adopted, so the carrier wavelength $\lambda$ is 0.125 m. Vehicles generate tasks, so source nodes and destination nodes are randomly selected from all vehicle nodes to avoid result bias caused by fixed paths. Through multiple simulations and parameter tuning, $C_{\mathrm{th}}$ is set to 1.5, and $\alpha,\beta,\gamma,\delta$ are set to 0.4, 0.2, 0.2, and 0.2, respectively. Other hyperparameter settings are shown in Table \ref{tab1}.

\begin{table}
	\begin{center}
		\renewcommand{\arraystretch}{1.5} 
		\caption{Hyperparameter settings}
		\label{tab1}
		\begin{tabular}{| c | c |}
			\hline
			\textbf{Parameter} & \textbf{Value}\\
			\hline
			Road length $L_{\mathrm{road}}$ & 20,000 m \\ 
			\hline 
			Road width $W_{\mathrm{road}}$ & 200 m \\
			\hline
			Channel bandwidth & 10 MHz \\
			\hline
			Transmit power $P_t$ & 20 dBm \\
			\hline
			Noise power spectral density  & $10^{-13}$ W/Hz \\
			\hline
			RSU communication range $R_r$ & 500 m \\
			\hline
			V2V maximum communication distance $R_v$ & 300 m \\
			\hline
			Reference speed threshold $v_{\mathrm{ref}}$ & 30 m/s \\
			\hline
			Minimum PDR threshold $\mathrm{PDR}_{\mathrm{min}}$ & 0.7 \\
			\hline
			Maximum end-to-end delay $T_{\mathrm{max}}$ & 10 s \\
			\hline
			Packet length $L_p$ & 10000 bit \\
			\hline
			$\sigma$, $\zeta$, $d_0$ & 8, 3.5, 1 \\
			\hline
		\end{tabular}
	\end{center}
\end{table}

To verify the performance superiority of the proposed scheme, four typical comparative algorithms in the field of IoV routing are selected, and the core design of each algorithm is as follows:
\begin{itemize}
	\item RSU-Prioritized V2V (RSU-V2V): Firstly, vehicles that are within the communication range of the same RSU as the source node and closer to the destination node are selected as candidates. When no available RSU exists, the algorithm degrades to traditional V2V direct communication. The next hop is determined by a static scoring function that combines transmission rate, link BER, and distance to the destination node.
	\item Load-Aware V2V (LA-V2V): Adopts full V2V direct communication. On the basis of the scoring dimensions of RSU-V2V, it adds the instantaneous load of nodes and avoids local congestion by penalizing highly loaded nodes. However, it only perceives node-level local load, without global congestion and link stability metrics.
	\item Model-based Reinforcement Learning (MRL): Constructs a Q-table with source-neighbor node pairs as state keys, calculates the immediate reward based on transmission rate, BER, and distance, and updates the Q-values accordingly. It adopts the  $\varepsilon$-greedy strategy, fusing the immediate reward and historical Q-values to select the next hop, thus achieving adaptive routing optimization.
	\item DRL for QoS Optimization (DRL-QoS): Constructs a four-dimensional state vector consisting of link interruption probability, node load, transmission rate, and BER, and designs an exponential reward function to minimize the deviation between QoS metrics and target values. It selects the next hop by combining reward values and distance through the $\varepsilon$-greedy strategy.
\end{itemize}

\begin{figure}[t]
	\centerline{\includegraphics[width=3.0in,keepaspectratio]{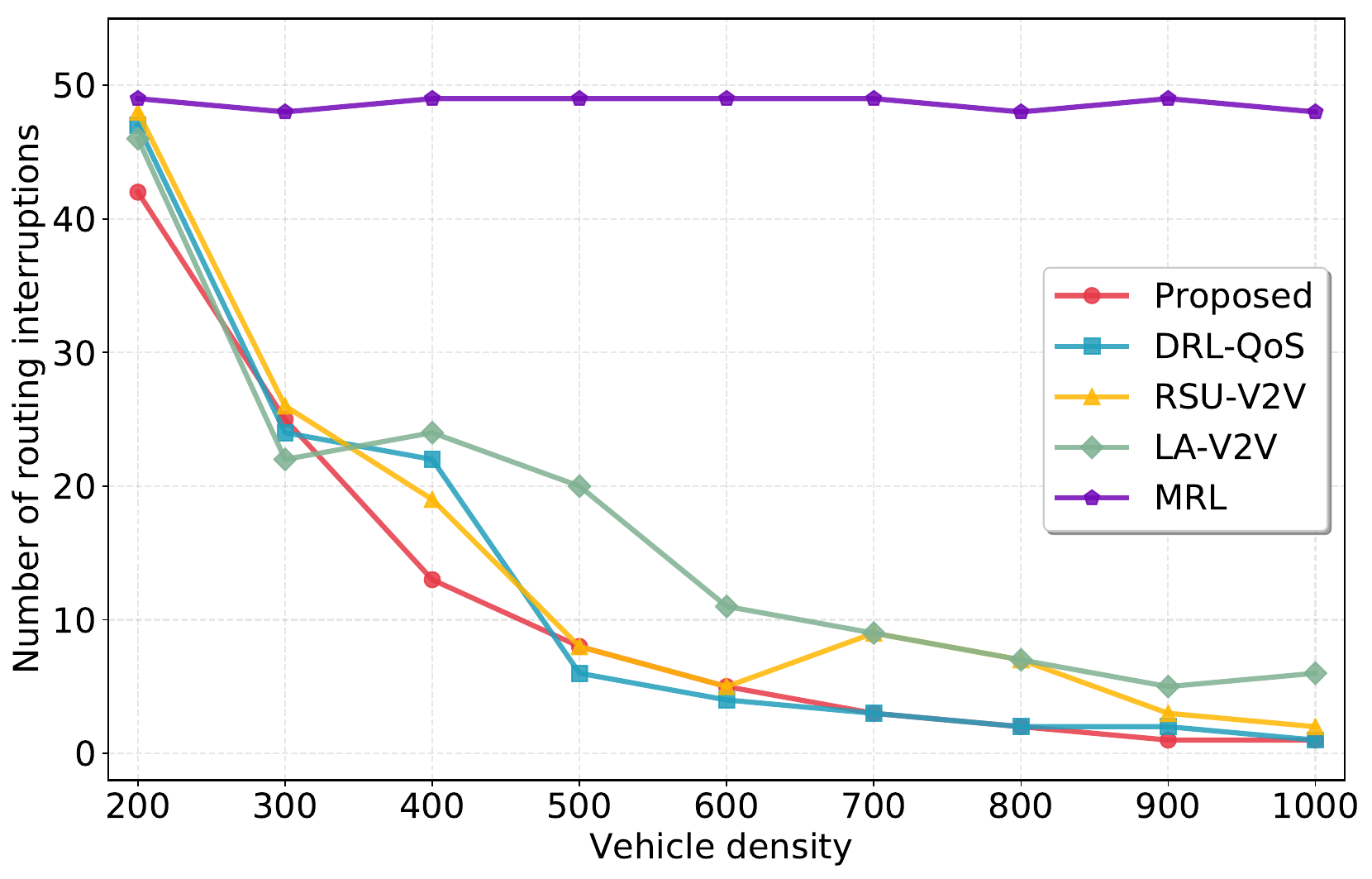}}
	\caption{The trend of routing interruption count versus vehicle density.}
	\label{fig2}
\end{figure}

Fig. \ref{fig2} illustrates the variation of routing disruption times with vehicle density. The proposed scheme exhibits a continuous decrease in disruption times from low to high vehicle density, outperforming the comparative algorithms significantly. This superiority verifies that the global congestion awareness, link stability metric, and primary-backup path switching mechanism of the proposed scheme can effectively adapt to the highly dynamic topology of IoV. Even under ultra-high-density congestion scenarios, the routing disruption rate can be maintained at an extremely low level. In contrast, LA-V2V only perceives local load, RSU-V2V lacks intelligent learning capabilities, and both DRL-QoS and MRL fail to incorporate global dimensions, rendering them unable to achieve comparable disruption control performance.

\begin{figure}[t]
	\centerline{\includegraphics[width=3.0in,keepaspectratio]{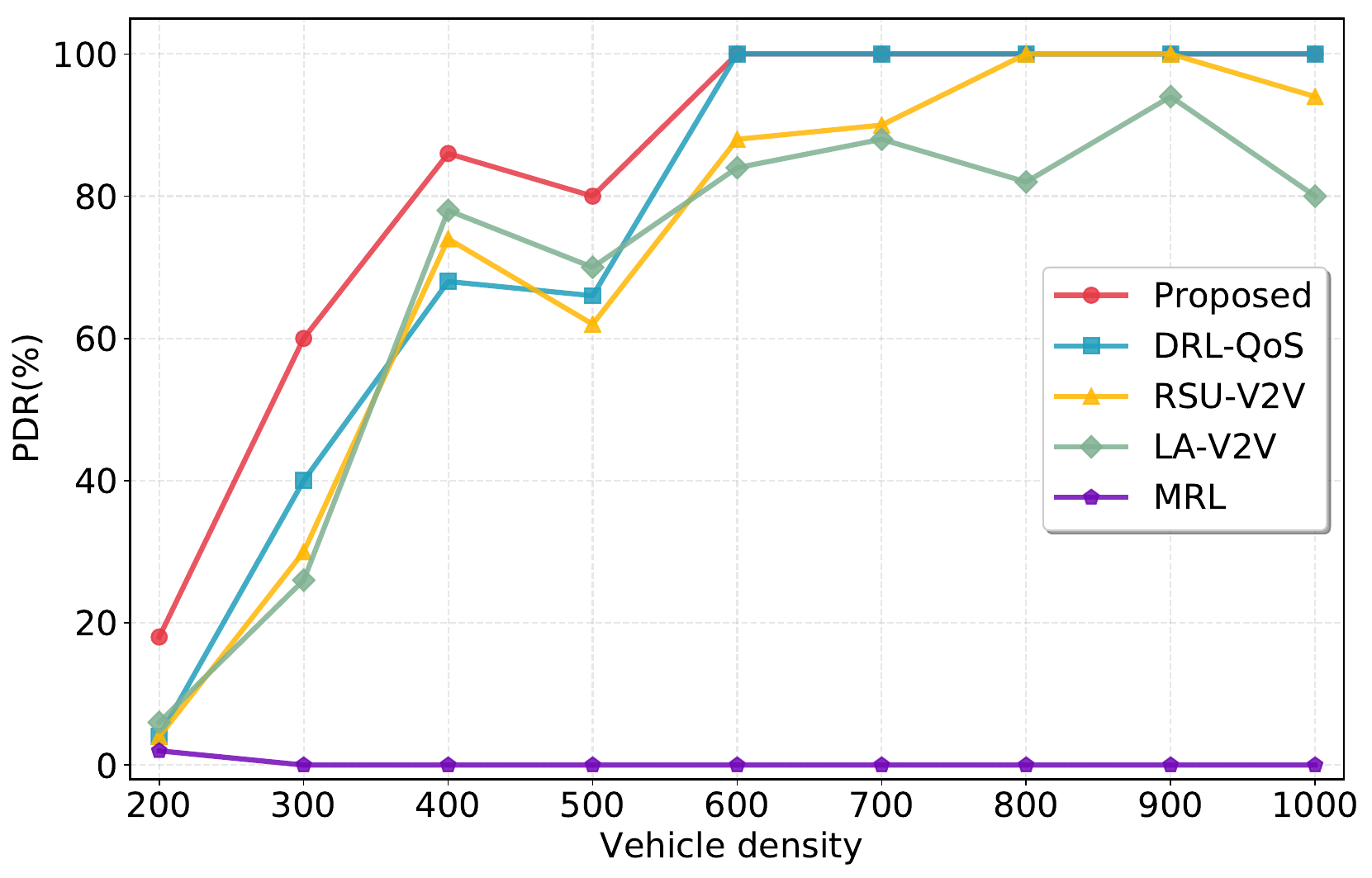}}
	\caption{The trend of PDR versus vehicle density.}
	\label{fig3}
\end{figure}

As shown in Fig. \ref{fig3}, the PDR of the proposed scheme is significantly improved with the increase of vehicle density, achieving favorable performance under various density scenarios and consistently outperforming the comparative algorithms. DRL-QoS and RSU-V2V also exhibit strong QoS guarantee capability in terms of PDR under high-density conditions. Limited by its single-node load perception only, LA-V2V achieves slightly inferior PDR performance at high density. Meanwhile, MRL suffers from the lack of congestion and stability awareness, resulting in a PDR close to 0, which cannot meet the communication requirements of IoV.
This indicates that the proposed scheme effectively enhances the packet delivery ratio via multi-metric evaluation and primary-backup path switching, and demonstrates outstanding robustness especially under heavy congestion scenarios.

\begin{figure}[t]
	\centerline{\includegraphics[width=3.0in,keepaspectratio]{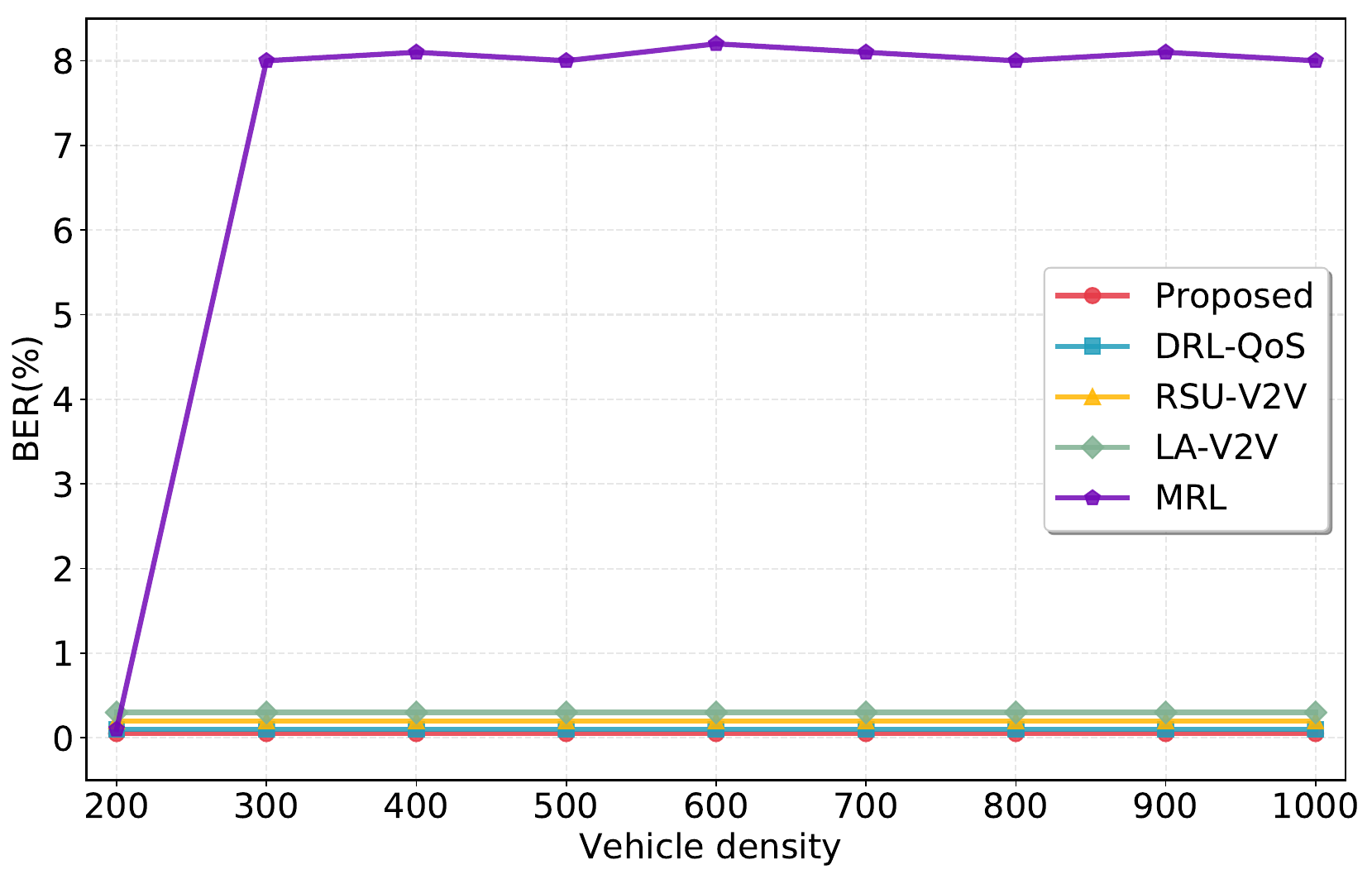}}
	\caption{The trend of BER versus vehicle density.}
	\label{fig4}
\end{figure}

\begin{figure}[t]
	\centerline{\includegraphics[width=3.0in,keepaspectratio]{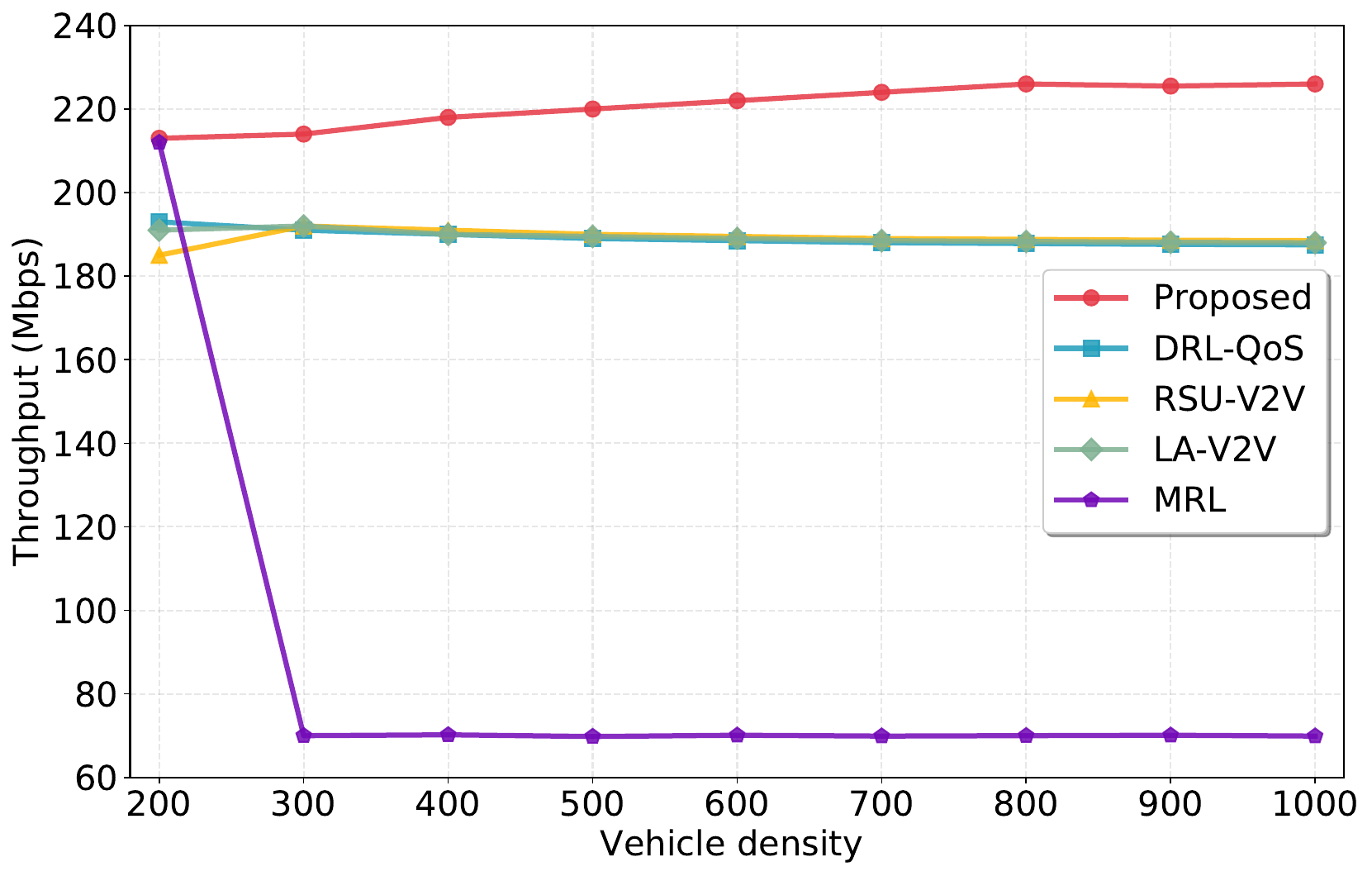}}
	\caption{The trend of throughput versus vehicle density.}
	\label{fig5}
\end{figure}

Link BER and throughput are key metrics for evaluating link quality and transmission efficiency of routing algorithms, which jointly determine the reliability and effectiveness of IoV communications. As shown in Figs. \ref{fig4} and \ref{fig5}, the proposed scheme exhibits significant advantages in both indicators. Its BER remains stably at an extremely low level, effectively avoiding high-error links; meanwhile, the throughput increases steadily with vehicle density, reaching approximately 226 Mbps in high-density scenarios, which is notably superior to the comparison algorithms. The BER and throughput of DRL-QoS, RSU-V2V, and LA-V2V are slightly inferior to those of the proposed scheme. In contrast, due to the lack of awareness of link quality, congestion, and stability, the MRL algorithm experiences a sharp rise in BER to around 8\% and a drastic drop in throughput to approximately 70 Mbps when vehicle density exceeds 300, failing to meet the core IoV requirements of low BER and high throughput.

\begin{figure}[t]
	\centerline{\includegraphics[width=3.0in,keepaspectratio]{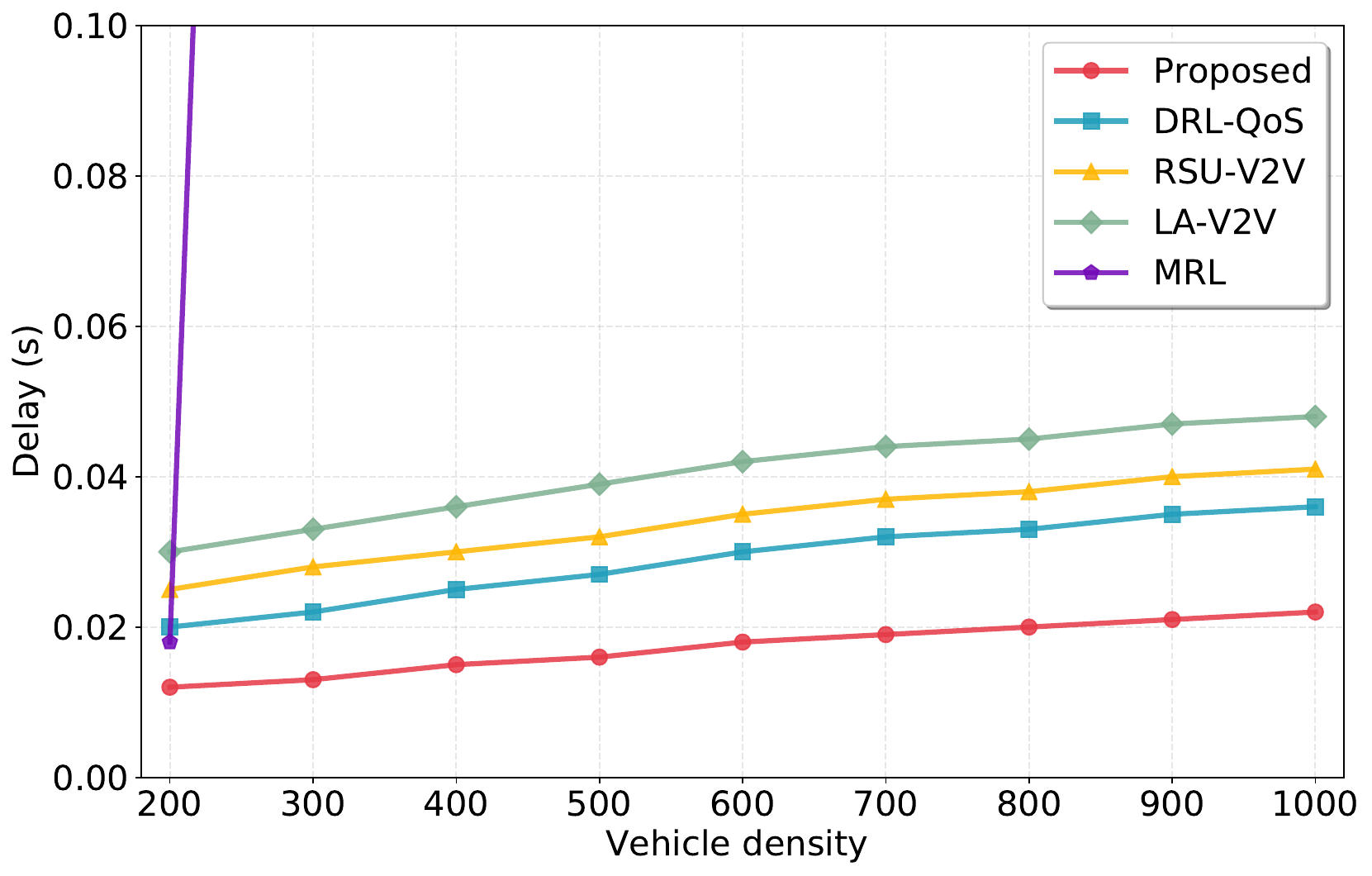}}
	\caption{The trend of delay versus vehicle density.}
	\label{fig6}
\end{figure}

As shown in Fig. \ref{fig6}, the end-to-end delay of the proposed scheme remains consistently the lowest and increases most gently with rising vehicle density, significantly outperforming the comparison algorithms. The delays of DRL-QoS, RSU-V2V, and LA-V2V rise slowly in the ranges of 0.020$\sim$0.036 s, 0.025$\sim$0.041 s, and 0.030$\sim$0.048 s, respectively, reflecting the impact of different sensing capabilities on latency control. In contrast, due to the lack of congestion and link stability awareness, the MRL algorithm exhibits generally high latency across all vehicle densities, failing entirely to meet the real-time requirements for IoV safety information transmission.

\begin{figure}[t]
	\centerline{\includegraphics[width=3.0in,keepaspectratio]{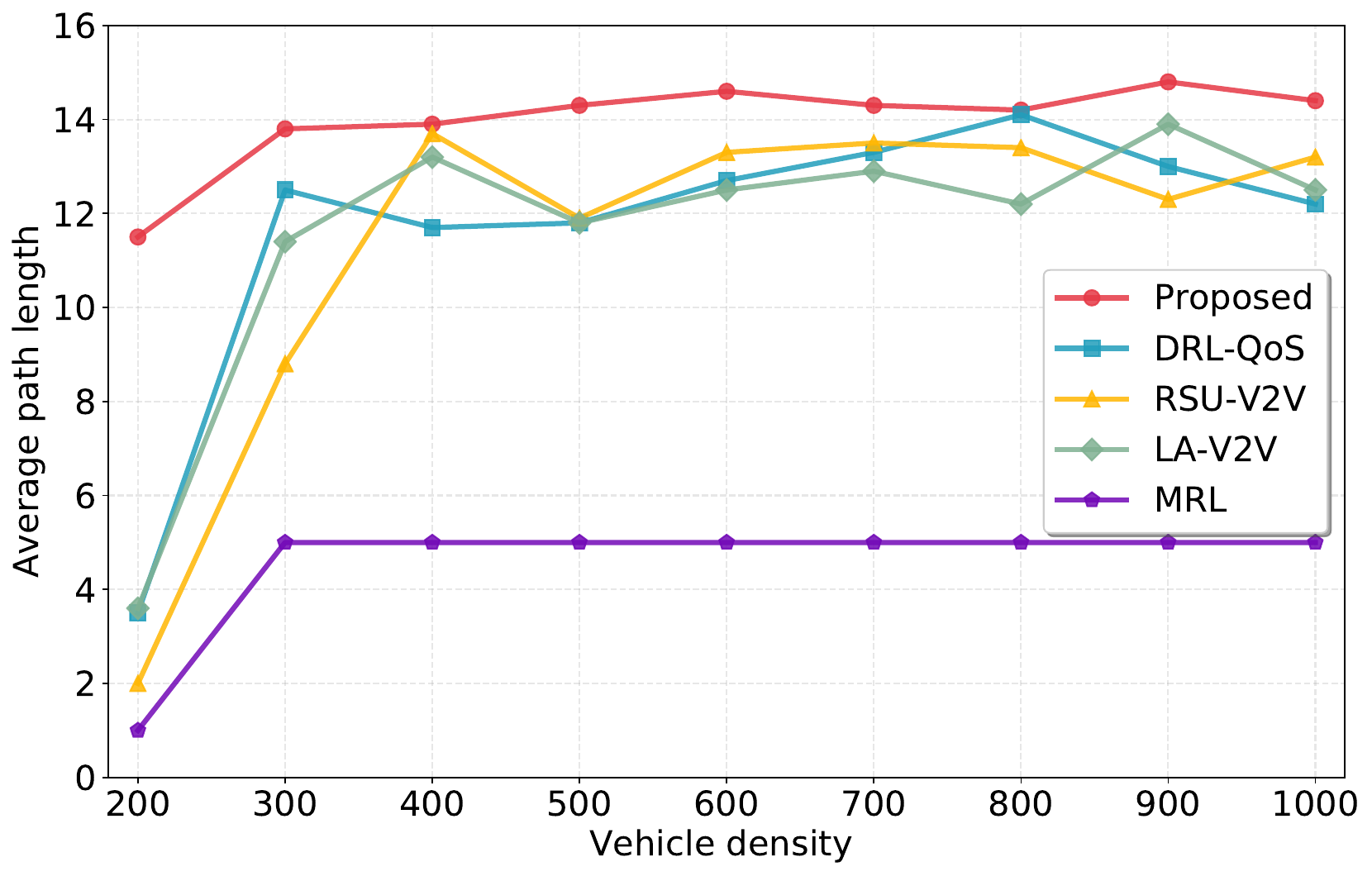}}
	\caption{The trend of average path length versus vehicle density.}
	\label{fig7}
\end{figure}

As shown in Fig. \ref{fig7}, the average path length of the proposed scheme increases steadily with vehicle density, reaching approximately 14.4 hops in high-density scenarios, which is significantly longer than that of the comparison algorithms. This is because the proposed scheme prioritizes paths with good link quality and low congestion rather than the shortest path in routing decisions, thereby effectively improving communication reliability. The average path lengths of RSU-V2V, LA-V2V, and DRL-QoS fluctuate between 11.7 and 13.9 hops, reflecting their trade-off between QoS guarantee and path length. In contrast, due to the lack of congestion and link quality awareness, the MRL algorithm always selects the shortest path, with an average path length stable at 5 hops, resulting in extremely poor communication reliability in high-density scenarios.

\begin{figure}[t]
	\centerline{\includegraphics[width=3.3in,keepaspectratio]{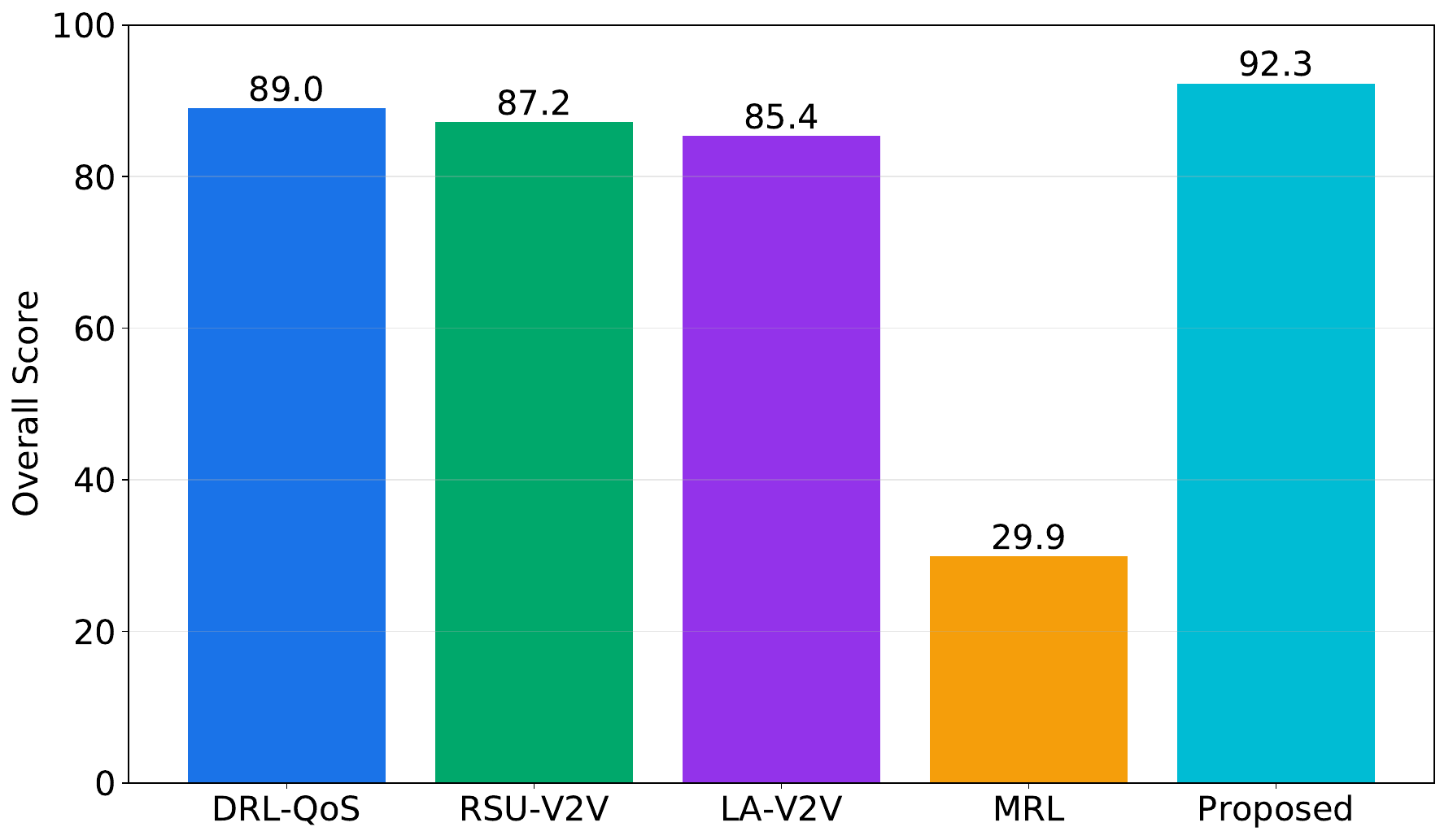}}
	\caption{Comprehensive Performance Scores of Different Algorithms.}
	\label{fig8}
\end{figure}

To comprehensively evaluate the overall performance of various algorithms in IoV scenarios, we weighted and normalized the number of interruptions, link BER, end-to-end delay, throughput, and PDR to obtain the overall scores shown in Fig. \ref{fig8}. The proposed scheme significantly outperforms others with a score of 92.3, while the scores of DRL-QoS, RSU-V2V, and LA-V2V are 89.0, 87.2, and 85.4, respectively, and the MRL algorithm achieves only 29.9. This result demonstrates that the proposed scheme achieves the optimal overall balance of communication reliability, real-time performance, and transmission efficiency in highly dynamic and congested IoV scenarios via its innovative multi-dimensional metric, primary-backup path switching, and V2I/V2V adaptive mechanism.

\section{Conclusion}\label{5}
To address issues such as high dynamic topology, channel fluctuation, and network congestion in IoV, this paper proposed an adaptive multi-dimensional coordinated comprehensive routing scheme for IoV. The scheme constructed a complete system model including network topology, communication links, hierarchical congestion, and transmission delay, defined a multi-dimensional routing metric fusing link reliability, node load, global congestion, and link stability, designed an intelligent V2I/V2V switching mechanism, combined primary-backup path dual decision-making and threshold switching strategy to avoid link interruption, and realized dynamic parameter adjustment through an adaptive function to adapt to network state changes. Simulation experiments comparing with four typical algorithms showed that the proposed scheme effectively reduced the number of routing interruptions and bit error rate, improved packet delivery rate and throughput, maintained low end-to-end delay with gentle growth, and achieved the global optimal balance of routing reliability, real-time performance, and transmission efficiency.

\bibliographystyle{ieeetr} 
\bibliography{MyRefs} 
~~~\\
~~~\\

\end{document}